\documentclass{article}
\usepackage{spconf,amsmath,graphicx}
\usepackage{color}
\usepackage{caption}
\usepackage{subcaption}
\usepackage{amsfonts}
\usepackage{bm}
\usepackage{blindtext}
\usepackage{svg}

\graphicspath{{figures/}}

\def\comment #1 {\textcolor{red}{#1}}
\def\rts{RT$_{60}$}

\title{Blind Acoustic Room Parameter Estimation Using Phase Features}

\name{$
\begin{array}{ccc}
	\mbox{Christopher Ick$^{1}$, Adib Mehrabi$^{2}$, Wenyu Jin$^{3}$}.
\end{array}
$}

\address{
$^{1}$ New York University, New York, NY 10003, United States \\
$^{2}$ Sonos Experience Limited, United Kingdom
$^{3}$ Sonos, Inc., Boston, MA 02111, United States
} 

\begin{document}
\ninept
\def\baselinestretch{.903}\let\normalsize\small\normalsize
\maketitle
\begin{abstract}

Modeling room acoustics in a real-world settings involves some degree of blind parameter estimation from noisy and reverberant audio.
Modern approaches leverage convolutional neural networks (CNNs) in tandem with time-frequency representations.
Using short-time Fourier transforms to develop these spectrogram-like features has shown promising results, but this method implicitly discards a significant amount of audio information in the phase domain.
Inspired by recent works in speech enhancement, we propose utilizing phase-related features to extend recent approaches to blindly estimate the so-called ``reverberation fingerprint'' parameters, namely, volume and RT{$_{60}$}.
The addition of these features is shown to outperform existing methods that rely solely on magnitude-based spectral features across a wide range of acoustics spaces. We evaluate the effectiveness of the deployment of these phase features in both single-parameter and multi-parameter estimation strategies, using a task-specific dataset that consists of publicly available room impulse responses (RIRs), synthesized RIRs, and in-house measurements of real acoustic spaces.

\end{abstract}



\section{Introduction}\label{sec:intro}
In recent years, the dynamic parameterization of listeners' local acoustic space has become increasingly relevant in the area of audio processing.
Knowledge of the local room acoustics can be used to enhance speech signals or aid dereverberation algorithms, for the purpose of word recognition or voice communication clarity \cite{Zhang18,Wu17,Mohammadiha16}.
Spatial sound reproduction systems \cite{Cecchi18,Jin16,Jin22} may also benefit from this information for acoustic room equalization tasks.
To this end, one of the major challenges is to estimate the best acoustical parameters of the room to improve the plausibility of immersive audio.
In \cite{Jot16}, the concept of "reverberation fingerprint", which consists of the room’s volume and its frequency-dependent diffuse reverberation decay time, was introduced to characterize rooms for realistic binaural rendering on audio augmented reality headphones.
This fingerprint is limited to the position-independent part of the reverberation as it characterizes a room in isolation from the orientation and directivity of sources and receivers.

Conventionally, room parameters such as reverberation time (\rts) and direct-to-reverberant ratio (DRR) can be derived directly from measured RIRs.
Room volume is highly related to the estimation of the so-called critical distance, which is defined as the distance at which the direct and reverberant power portions of a sound source become equivalent, i.e., the point at which the DRR is 0 dB.
Under the ideal diffuse soundfield assumption, the relation between these parameters is given by the well know Sabine’s equation \cite{Kuttruff16}.
In practice, in-situ RIR measurements of users’ local acoustic environments are typically difficult to achieve. An attractive alternative is to estimate room acoustic parameters blindly from audio recordings of unknown sound sources using microphones in the presence of noise.
The 2015 ACE challenge \cite{Eaton16} set the bar for blindly estimating \rts\, and DRR from noisy speech signals in the presence of various noise types, with signal-modelling based approaches for \rts  \cite{Prego15,Loellmann15} being the top-performing systems. On the other hand, room volume estimation has long been formulated as a classification problem.
The systems in \cite{Moore14,Peters12} leverage MFCC-based features to identify the room that corresponds to a environmental sound or speech recording from a closed set for the purpose of audio-forensics.
In \cite{Murgai17}, a blind estimation algorithm for the reverberation fingerprint was proposed based on the decay envelope of a single-channel clean speech signal using simulated data.

Recent advancements in deep neural networks also push the boundary of room acoustic parameters estimation.
An approach of blind room volume estimation from single-channel noisy speech deploying a CNN was proposed in \cite{Genovese19}, estimating volume within a factor of 2 from the ACE challenge \cite{Eaton16}.
A number of neural-network-based systems were proposed to blindly estimate \rts\, and DRR from single-channel speech signals \cite{Xiong15,Gamper18,Bryan20,Gotz22} and demonstrated promising results in terms of both estimation accuracy and the ability to follow temporal variations in dynamically evolving acoustic environments.
Srivastava et al. \cite{Srivastava21} devised a system that jointly estimates multiple room parameters including the total surface area, the volume, the frequency-dependent \rts\, and mean surface absorption coefficients. However, the system relies on two-channel noisy speech recordings to explore inter-channel features, which implies it is more hardware-demanding.

In this paper, we propose using a CNN model to estimate room acoustic parameters blindly from single-channel noisy speech signals with the special focus on the reverberation fingerprint (i.e. volume and \rts).
Unlike prior works that generally rely on the log-energy calculation of spectro-temporal features, we introduce phase-related features.
To the authors' best knowledge, this is the first DNN-based room parameter estimation system that leverages phase-related features. 
Specifically, the Gammatone phase spectrogram and its numerical derivatives that denote relative phase variations across the time and frequency axises are considered. Results on both simulated and real-world RIRs show that the proposed model outperforms state-of-the-art single channel blind room acoustic parameter estimation systems on similarly-structured datasets.

\section{Data generation pipeline}

Approaching the task of room parameter estimation with neural-network-based approaches is inherently challenging due to the scale of data required.
Because the task relies on having audio samples from a wide variety of rooms, building an appropriately diverse dataset by hand would be impractically expensive and time consuming.
For our work, we instead develop a multi-stage audio generation pipeline that makes use of a wide range of RIRs across various room configurations with labeled volume and broad-band \rts.

\subsection{Real-world RIR Dataset}

To maximize the range of room acoustic parameters, we sampled audio simulations from four different publicly available datasets to provide 43 distinct room profiles, with volumes ranging from single-person phone booths, to nuclear reactor halls.
Meanwhile, \rts  values ranging from less than half a second, to several over 10 seconds, which were computed using the Schroeder method \cite{schroeder}.
To ensure consistency across datasets, all RIRs were downsampled to 16 kHz.
Because many of these datasets were captured in academic settings, most rooms are rectangular in shape, and fall generally into the category of offices, classrooms, and auditoriums/lecture halls.
This includes the ACE Challenge dataset \cite{Eaton16}, the Aachen impulse response (AIR) dataset \cite{jeubair}, and the Brno University of Technology Reverb Database (BUT ReverbDB) \cite{butreverb}.
Lastly, the OpenAIR dataset \cite{murphy2010openair:} covers primarily large acoustic spaces (cathedrals, nuclear reactors, and other large structures). 
To account for the natural gap of real-world acoustic spaces falling in the 100 m{$^3$} to 1000m{$^3$} range of volumes, we made several in-house RIR measurements within this range to augment our available data, by recording sine-sweeps in shoebox-shaped rooms of measured volume.

\subsection{Synthetic RIRs}
\begin{figure}
    \centering
    \includegraphics[width=7cm]{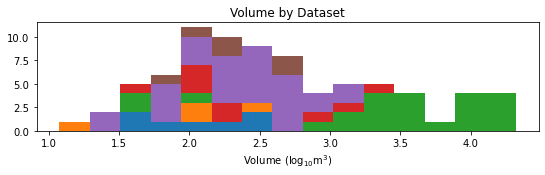}
    \includegraphics[width=7cm]{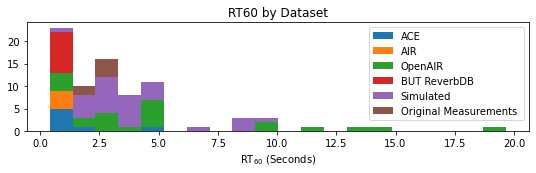}
    \caption{A histogram showing the distribution of RIR over various volumes and \rts  \, ranges from various datasets}
    \label{fig:data_distribution}
\end{figure}

To increase the scale of data, particularly in the range of less frequently available room volumes, we added in 30 simulated RIRs for rooms of various geometries.
Each room was simulated with a singular source near the center of the room, and five point receivers uniformly sampled over the volume of each room.
To generate this dataset, we made use of \textit{pyroomacoustics} \cite{pyroomacoustics}, a software package that uses the image-source model for RIR simulation for specific room geometries.
While this geometric method doesn't account for effects such as diffraction and scattering, experiments show that the use of simulated data benefits the performance of the model, which generalizes well to natural data \cite{Genovese19}. Overall, Fig. \ref{fig:data_distribution} illustrates the distribution of  room volumes and \rts on logarithmic and linear scales, respectively.
To ensure that our data was not overfitting to the well-behaved simulated data, our validation and test data splits did not include any data from the simulated datasets (except in the case of experiments that were explicitly measuring the performance of the model on simulated data).

\section{Audio Generation}

From this RIR dataset with known volume and \rts\, values, we could then generate reverberant audio data that would be eventually fed into our CNNs.
To do this, we convolved speech data recorded in anechoic chambers with RIRs, effectively mapping the acoustic response of the room to the audio.
This audio was combined with white noise convolved with the same RIR, the same method for noisey data generation as done in the ACE challenge \cite{Eaton16}.
To enforce robustness across various noisiness conditions, we combined these two signals to simulate the recording at 5 distinct SNR levels, including no noise (+Inf) and [+30,+20,+10, +0]dB.

Speech samples were drawn from the ACE dataset \cite{Eaton16} of anechoic voice recordings in randomly selected 4 second increments, with both male and female speakers. Within the RIR dataset, certain rooms had more RIR measurements than others, so to ensure uniformity of representation, each room was equally sampled, so that the distribution of audio samples in our dataset matched the distribution of volumes in our RIR dataset, as seen in Fig. \ref{fig:data_distribution}.

The data was split into a 6-2-2 train, validation and test split by rooms, as shown in Table 1. Specific rooms were selected for the validation and test splits to ensure a wide range of specific volumes that remained unseen by the models during the training.

\begin{table}
\begin{tabular}{c|ccc}
Data Split & \# of  & Real Rooms & Simulated Rooms\\
\hline
Train & 12000 & 29 & 18 \\
Validation & 4000 & 16 & 12 \\
Test & 4000 & 16 & 0 \\
\end{tabular}
\caption{Summary of data splits}
\label{table:data-splits}
\end{table}

\section{Featurization}

As the primary input to a CNN, the featurization of audio into a two-dimensional time-frequency representation has a non-trivial amount of influence on the performance of the model.
Multiple featurization schemes can be combined to train a single model, but increasing the dimensionality of the input space can lead to highly complex models that are prohibitively data hungry and costly to train.
As such, choosing an audio representation that maximally captures information about the acoustic space must be balanced by discarding features that provide no benefit to the model.
This ensures that we are not confined by the relatively small data size we have access to, as well as hopefully providing more generalizable results.

\subsection{Spectral Representations}
Prior literature \cite{Genovese19,Gamper18,Srivastava22} suggests the importance of low-frequency effects for room acoustic parameter estimation, constraining their feature representations to relatively low frequency bands ($<2000$ kHz).
To keep model complexity low while preserving signal information that is relevant for the parameter estimation problem, we used Gammatone ERB filterbank to generate our time-frequency representation.
The Gammatone filterbank has 20 bands, ranging from 50Hz to 2000Hz.
In computing the STFT of our audio, we used a Hann window size of 64 samples, with a hop size of 32. The resulting spectrogram is convolved with our filterbank, resulting in a 20 $\times$ 1997 complex Gammatone spectrogram.
Additionally, as detailed in \cite{Genovese19}, a number of separate energy-based features that aim to capture the low-frequency behaviour were considered specifically for improving the room volume estimation performance.

\subsection{Phase Representation}
During the calculation of the log-energy of the Gammatone spectrogram, phase information is typically discarded.
However, recent literature in the area of speech enhancement has demonstrated the effectiveness of including various forms of phase information, either explicitly or implicitly \cite{kimphase}.
Being inspired by this, we propose retaining the phase information captured in the STFT and using it as an input for similar networks.
By computing the phase angle at every time-frequency bin of Gammatone features, we can generate a Gammatone phase spectrogram equivalently sized to our log-energy spectrogram.

In addition to the direct phase calculations, recent work in \cite{kimphase} has shown value in numerical derivatives of the phase spectrograms.
This provides us with a new feature that could potentially better characterize room reflection patterns at low frequencies, compensating for the nature of low spectral resolution due to the ERB filterbank.
Additionally, we also compute the \textit{phase continuity} $\phi$, which is a second degree partial derivative of the phase information along the frequency axis $k$, and the time axis $n$.
\begin{equation}
    \phi(\theta^k_n) = \lim_{i \rightarrow 0}\lim_{j \rightarrow 0}\frac{f(\theta^k_n)-f(\theta^{k+i}_{n+j})}{\theta^k_n-\theta^{k+i}_{n+j}}
\end{equation}
where $f(\theta) = \cos(\theta) + \sin(\theta)$ is a phase-wrapping function as defined in \cite{kimphase}. By concatenating various features together, we can compare the performance benefits of these features, while observing the trade-offs of training time and model complexity.

\section{Experiment Design}

\subsection{Network Architecture}
\begin{figure*}
    \centering
    \includegraphics[width=\textwidth, height=2.5cm]{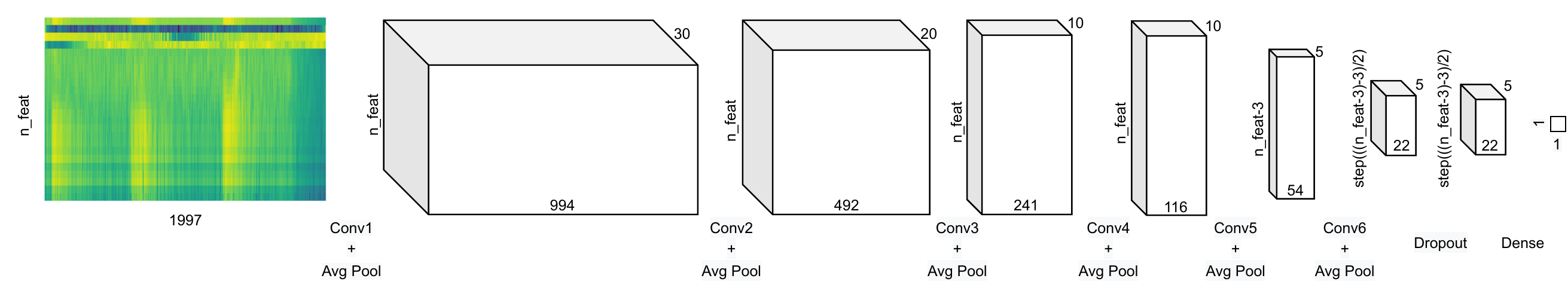}
    \caption{A visualization of our CNN architecture from our featurization. Note that the height of each layer is dependent on the dimensionality of the input feature, which varies from experiment to experiment}
    \label{fig:CNN-arch}
\end{figure*}

The model architecture was selected to be a feed-forward 2D CNN due to its suitability for capturing 2-dimensional time-frequency signal patterns.
Because the focus of this study is comparison of audio featurizations, the choice of architecture was not critical, and such, this relatively simple architecture was employed, for it's success in the same task in \cite{Genovese19}.
The model makes use of 4 time-domain convolutional layers, followed by 2 spectral convolutions, resulting in a total of 6 convolutional layers, each one containing an average pooling layer.
The convolutional block was then followed by a 50\% dropout layer and a final, single fully-connected layer that mapped down to output dimensionality.

Fig. 2 illustrates the full architecture employed.

\subsection{Evaluation Metrics}
The models were evaluated on their ability to estimate the room parameters within a feasible accuracy.
Perceptual evaluation of the dataset had shown that there were many more variations in the lower volume range than across the wide range of larger volumes, so to account for that, we evaluated our data on the base-10 logarithm of the volume.
This also mitigated the effect that larger acoustic spaces would be over-weighted in training due to the relative higher contribution to error estimations.

While this effect is less pronounced for \rts, we also computed the log of \rts for evaluation purposes, as it allows smaller values closer to 0 to be more widely distributed.

We evaluated the model on several standard regression metrics, including mean squared error (MSE) and the Pearson correlation coefficient ($\rho$) between our target and predicted values.
One additional metric we used is the exponential of the mean absolute logarithm of the ratio between estimated and target room parameters. Given target $y$, and estimate $\hat{y}$, we can write this as:

\begin{equation}
    \text{MeanMult} = \text{exp}\left(\frac{1}{N} \sum_{n=1}^N \left| \text{ln} \left( \frac{\hat{y}_n}{y_n}\right) \right| \right)
\end{equation}

This metric summarizes mean error in terms of the ratio between estimated and target parameters in a method that is less sensitive to overall scale, allowing more equal weighting across all scales.

Finally, we used one additional metric comparing the variance between all estimates and all targets, which is mainly used to determine if our model is overfitting to a singular estimation, in which case it approaches 0, or is capturing the range of our dataset, in which case it approaches 1.
This can be written as $VR=\frac{\text{var}(\hat{y}_n)}{\text{var}(y_n)}$.

\subsection{Training}
Each model was trained using MSE on the log of the target parameter using Pytorch's Adam optimizer.
Each training run ran for 1000 epochs, used L2 regularization, and implemented an adaptive learning rate to ensure efficient convergence.
We used grid search to tune hyperparameters including batch size, initial learning rate, regularization strength.
We varied the feature combinations on each model as inputs, but left the optimized hyperparameters the same to ensure consistency across model configurations, so that variations in performance would be due to the variations in feature representation.

\section{Results}
\subsection{Room parameter estimation using phase-related features}
To assess the effectiveness of phase-related features, we trained several models containing a variety of feature combinations.
As our baseline, we used a model trained on log-scaled magnitude spectral features for \rts estimation \cite{Gamper18}, as well as a set of additional hand-picked features that were proposed in \cite{Genovese19} for volume estimation.

Based on our early exploratory results, we built a phase-augmented model (\textit{+Phase}), taking the same log-scaled magnitude spectral coefficients, and concatenating them with the phase spectrogram features and the first order derivative of the phase, as described in Sec. 4.2.
To limit the dimensionality of our input features, these phase features were truncated to only include the bands corresponding to frequencies at, or below 500Hz, as these lower frequencies typically carry more information corresponding to the acoustic parameters of interest.
We found that this model best leveraged the additional phase information provided by phase spectrograms and it's first-order derivatives while also balancing model complexity for volume estimation.
Similarly, we also built an additional model (\textit{+Continuity}) by adding the continuity features described in \cite{kimphase} to our \textit{+Phase} model limited to sub-500Hz frequencies. Each of these models were trained and evaluated on the same data splits, and contained the same model architecture other than the input layer.
\begin{table}
\centering
Volume Estimation
\vspace{1mm}

\begin{tabular}{r|ccccc}
Feature Set & MSE  & $\rho$-corr.& \textit{MM} & \textit{VR}\\
\hline
Baseline \cite{Genovese19} & 0.4184  & 0.5736 & 1.6941 & 0.6624 \\
\textit{+Phase} & \textbf{0.3630}  & \textbf{0.6479} &  \textbf{1.6084} & \textbf{0.7461}  \\
\textit{+Continuity} & 0.3843 & 0.6299 & 1.6365 & 0.7415 \\
\end{tabular}

\vspace{1mm}
\rts \,Estimation
\vspace{1mm}
\begin{tabular}{r|ccccc}
Feature Set & MSE  & $\rho$-corr.& \textit{MM} & \textit{VR}\\
\hline
Baseline \cite{Gamper18} & 0.1754  & 0.9159 & 2.0424 & \textbf{0.7928} \\
\textit{+Phase} & 0.1569  & 0.9330 &  1.9687 & 0.7842  \\
\textit{+Continuity} & \textbf{0.1408} & \textbf{0.9463} & \textbf{1.9210} & 0.7840\\
\end{tabular}
\caption{Performance of volume and \rts\, estimation across baseline (log-magnitude), phase augmented, and continuity augmented models}
\label{table:exp1-results}
\end{table}

Table \ref{table:exp1-results} shows the performance comparison between investigated models. As it can be seen, in general, the two phase-relevant models feature clear performance benefits over the baseline, despite being slightly more complex due to a higher input dimensionality.
In terms of volume estimation, \textit{+Phase} model demonstrates the best performance,  achieving the lowest MSE as well as the highest Pearson correlation coefficient.
The inclusion of additional phase-continuity information in the \textit{+Continuity} does not lead to improved performance for the volume estimation task, perhaps due to the additional parameterization requiring more data and processing power to fully take advantage of.
However, the additional phase continuity features do provide improved performance over the baseline model as well as the basic phase-augmented model in the estimation of \rts, which confirms that the tracking of phase variations across time better characterizes \rts\, behaviors.

\begin{figure}
    \centering
    \includegraphics[height=2.86cm]{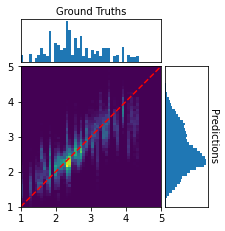}
    \includegraphics[height=2.86cm]{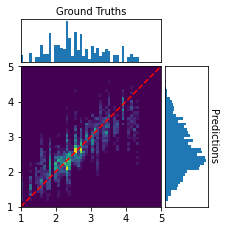}
    \includegraphics[height=2.86cm]{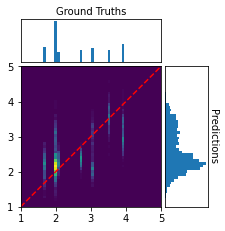}
    \caption{Confusion Matrices of our volume estimation model using \textit{+Phase} features on our train, validation, and testing dataset splits (from left to right).The dashed red line indicates a perfect prediction.}
    \label{fig:confusion-matrices}
    \vspace{-3mm}
\end{figure}

Fig. \ref{fig:confusion-matrices} visualizes the confusion matrices of the proposed volume estimation model for the training, the validation and the test set with x and y axis indicating index numbers of base-10 logarithm values of volume sizes.
The volume estimates are well distributed around the ground truth across the tested range, indicating that the CNN successfully learned a representation of the underlying regression problem.
Note that the variance is generally smallest in the 100 m$^3$ to 1000 m$^3$ range of estimations, where the bulk of our data lies  (including both real and simulated RIRs as shown in Fig. 1); estimation variance is more challenging outside this range due to both a lack of data, as well as the diverging effects of reverb in small acoustic spaces (which are relatively minimal) and large ones (which can be exceptionally hard to process due to long \rts\, values).
 
It's important to note that the model architecture is tuned to support the baseline featurization as is done in \cite{Genovese19}; this additionally complex featurization could be better leveraged by a deeper model with wider filters, or a multi-channel featurization as shown in sound event detection literature \cite{ick}.

\subsection{Joint estimation of room parameters}
The purpose of this section is to investigate the feasibility of jointly estimating multiple room parameters with similar performance to training separate parallel models. The assumption is that parameters such as \rts\, and volume draw reverberation information from similar frequency bands and acoustic characteristics, so that we can optimize the simultaneous estimation of all parameters jointly by training a single model. Note that we limited to \rts\, and volume due to the lack of ground truth for other room parameters (e.g. total surface area and mean surface absorption coefficients) in the considered real-world RIR datasets.

The architecture remained largely the same as the \textit{+Phase} and \textit{+Contiuity} in the prior section, except we modified the final output layers in both models to fully-connected layers that output two parameters (instead of only one). To avoid issues due to significant scale differences between room volume in m$^3$ and \rts in seconds, both (log-scaled) parameters were normalized such that the maximum value in each dataset was 1; we then trained the models on the sum of the normalized MSE for each parameter to ensure equal weighting. Table 3 demonstrates correlation coefficient scores for different systems. Note that other evaluation metrics are not selected for the purpose of consistency due to that parameters are scaled differently in the normalization process. Compared to the results in Table \ref{table:exp1-results}, we can observe very similar performance between the joint estimation systems and the same architectures that estimate each of these quantities individually with minimum performance loss. In particular, the joint system with the feature set of \textit{+Continuity} provides the most stable prediction when estimating volume and \rts\, simultaneously.
\begin{table}
\centering
Pearson correlation coefficient ($\rho$)
\begin{tabular}{c|cc}
Method & \rts  &  Volume\\
\hline
Baseline [16] &-  & 0.5736 \\
Baseline [18] & 0.9159  & -  \\
\textit{+Phase} w/joint & 0.9304  & 0.6090 \\
\textit{+Continuity} w/joint & \textbf{0.9445}  & \textbf{0.6195}  \\
\end{tabular}
\vspace{1mm}
\caption{Pearson correlation coefficients of parameter estimates using various systems.}
\label{table:exp2-results}
 \vspace{-3mm}
\end{table}

\section{Conclusion}

In this paper, we propose improving the task of room parameter estimation by utilizing phase-related features, including phase spectrograms, numerical derivatives of phase, and phase continuity.
We demonstrate the benefit of these features in the context of blind parameter estimation from single-channel noisy speech recordings using a CNN architecture.
We assemble a dataset from a variety of sources, including publicly available RIR datasets, acoustic room simulations, and in-house RIR measurements.
Using the proposed phase-related features, we're able to demonstrate clear improvements over previous state-of-the-art methods in terms of reverberation fingerprint estimation on unseen real-world rooms.
Furthermore, we demonstrate the effectiveness of the proposed method in a joint-estimation configuration, showing that sharing parameters improves stability over single-parameter models.

The use of phase-related features for parameter estimation should potentially be extended to multi-band \rts, absorption coefficients, or other acoustic parameters, and is an area for future work.
Further opportunity  lies in additional, more-complex phase-related featurizations; recent work in sound event detection demonstrate benefits from multichannel approaches to spectral featurizations \cite{ick} allowing for cross-feature representations through 3-D convolution, which could prove beneficial in this task.

\newpage

\bibliographystyle{IEEEbib}
\bibliography{refs}

\end{document}